%
%
%
%
%
\magnification=\magstep1
\def\pn{\par\noindent}
%
%
\def\e{\hbox{\rm e}}
\def\m@th{\mathsurround=0pt}

\def\fsquare(#1,#2){
\hbox{\vrule$\hskip-0.4pt\vcenter to #1{\normalbaselines\m@th
\hrule\vfil\hbox to #1{\hfill$\scriptstyle #2$\hfill}\vfil\hrule}$\hskip-0.4pt
\vrule}}

\def\addsquare(#1,#2){\hbox{$
        \dimen1=#1 \advance\dimen1 by -0.8pt
        \vcenter to #1{\hrule height0.4pt depth0.0pt%
        \hbox to #1{%
        \vbox to \dimen1{\vss%
        \hbox to \dimen1{\hss$\scriptstyle~#2~$\hss}%
        \vss}%
        \vrule width0.4pt}%
        \hrule height0.4pt depth0.0pt}$}}

\def\Addsquare(#1,#2){\hbox{$
        \dimen1=#1 \advance\dimen1 by -0.8pt
        \vcenter to #1{\hrule height0.4pt depth0.0pt%
        \hbox to #1{%
        \vbox to \dimen1{\vss%
        \hbox to \dimen1{\hss$~#2~$\hss}%
        \vss}%
        \vrule width0.4pt}%
        \hrule height0.4pt depth0.0pt}$}}

\def\Fsquare(#1,#2){
\hbox{\vrule$\hskip-0.4pt\vcenter to #1{\normalbaselines\m@th
\hrule\vfil\hbox to #1{\hfill$#2$\hfill}\vfil\hrule}$\hskip-0.4pt
\vrule}}

\def\naga{%
        \hbox{$\vcenter to 0.4cm{\normalbaselines\m@th
        \hrule\vfil\hbox to 1.2cm{\hfill$\cdots$\hfill}\vfil\hrule}$}}

\def\Frect(#1,#2,#3){
\hbox{\vrule$\hskip-0.4pt\vcenter to #1{\normalbaselines\m@th
\hrule\vfil\hbox to #2{\hfill$#3$\hfill}\vfil\hrule}$\hskip-0.4pt
\vrule}}

\def\PFrect(#1,#2,#3){
\hbox{$\hskip-0.4pt\vcenter to #1{\normalbaselines\m@th
\vfil\hbox to #2{\hfill$#3$\hfill}\vfil}$\hskip-0.4pt}}

\dimen1=0.5cm\advance\dimen1 by -0.8pt

\def\vnaka{\normalbaselines\m@th\baselineskip0pt\offinterlineskip%
        \vrule\vbox to 0.6cm{\vskip0.5pt\hbox to \dimen1{$\hfil\vdots\hfil$}
       \vfil}\vrule}

\dimen2=1.5cm\advance\dimen2 by -0.8pt

\def\vnakal{\normalbaselines\m@th\baselineskip0pt\offinterlineskip%
        \vrule\vbox to 1.2cm{\vskip7pt\hbox to \dimen2{$\hfil\vdots\hfil$}
\vfil}\vrule}

%
%
\def\R{(2.1)}
\def\Ham{(2.2)}
\def\Rmod{(2.3)}
\def\trotter{(2.4)}
\def\staggered{(2.5)}
\def\free{(2.6)}
\def\corrlength{(2.7)}
\def\embedding{(2.8)}

\def\boxo{(3.1)}
\def\eig{(3.2)}
\def\baeo{(3.3)}
\def\prodbox{(3.4)}
\def\lambox{(3.5)}
\def\condo{(3.6)}
\def\funco{(3.7)}

\def\hdef{(4.1.a)}
\def\ndef{(4.1.b)}

\def\lamamtj{(4.2.a)}
\def\lam0mtj{(4.2.b)}
\def\lama0tj{(4.2.c)}

\def\freltjo{(4.3.a)}
\def\freltjtw{(4.3.b)}
\def\freltjth{(4.3.c)}
\def\freltjfo{(4.3.d)}

\def\ytranstjo{(4.4.a)}
\def\ytranstjtw{(4.4.b)}

\def\ysys{(4.5)}
\def\ysystjo{(4.5.a)}
\def\ysystjtw{(4.5.b)}
\def\ysystjth{(4.5.c)}
\def\ysystjfo{(4.5.d)}
\def\ysystjfi{(4.5.e)}

\def\ytilde{(4.6)}

\def\ninttjo{(4.7.a)}
\def\ninttjtw{(4.7.b)}
\def\ninttjth{(4.7.c)}
\def\ninttjfo{(4.7.d)}
\def\ninttjfi{(4.7.e)}
\def\ninttjthd{(4.7.c')}

\def\freetj{(4.8)}

\def\chgo{(4.9.a)}
\def\chgtw{(4.9.b)}
\def\chgth{(4.9.c)}

\def\scho{(4.10.a)}
\def\schtw{(4.10.b)}
\def\schth{(4.10.c)}
\def\schfo{(4.10.d)}

\def\frelseho{(5.1.a)}
\def\frelsehtw{(5.1.b)}
\def\frelsehth{(5.1.c)}
\def\frelsehfo{(5.1.d)}
\def\frelsehfi{(5.1.e)}

\def\eqalityseh{(5.2)}

\def\ytranseho{(5.3.a)}
\def\ytransehtw{(5.3.b)}
\def\ytransehth{(5.3.c)}

\def\ysysseho{(5.4.a)}
\def\ysyssehtw{(5.4.b)}
\def\ysyssehth{(5.4.c)}
\def\ysyssehfo{(5.4.d)}
\def\ysyssehfi{(5.4.e)}
\def\ysyssehsi{(5.4.f)}

\def\nintseho{(5.5.a)}
\def\nintsehtw{(5.5.b)}
\def\nintsehth{(5.5.c)}
\def\nintsehfo{(5.5.d)}
\def\nintsehfi{(5.5.e)}
\def\nintsehsi{(5.5.f)}

\def\eigseh{(5.6)}

\def\chgseh{(5.7)}

\def\excyzero{(6.1)}
\def\excrenorm{(6.2)}

\def\excTBAo{(6.3.a)}
\def\excTBAtw{(6.3.b)}
\def\excTBAth{(6.3.c)}

\def\condxm{(6.4)}
%
%
\def\YY{1}
\def\Gaud{2}
\def\HeisenTak{3}
\def\TakSuz{4}
\def\Vla{5}
\def\KluZitt{6} 
\def\EKSstring{7}
\def\JuDo{8}
\def\spinkb{9}
\def\spinkbt{10}
\def\AlczMart{11}
\def\AlZ{12}
\def\TsvBab{13}
\def\MSuz{14}
\def\InSuz{15}
\def\Koma{16}
\def\SAW{17}
\def\SNW{18}
\def\TakQT{19}
\def\Klu{20}
\def\KZeit{21}
\def\DdV{22}
\def\Mizu{23}
\def\KWZ{24}
\def\JK{25}
\def\JKSone{26} 
\def\JKStwo{27} 
\def\Cherednik{28}
\def\CPo{29}
\def\CPt{30}
\def\KNS{31}
\def\KN{32}
\def\Martins{33}
\def\Fendley{34}
\def\KM{35}
\def\KNSCh{36}
\def\TateoDorey{37}
\def\BLZ{38}
\def\Tsuboi{39}
\def\Schlottmann{40}
\def\EK{41}
\def\EKS{42}
\def\SuthTJ{43}
\def\BB{44}
\def\BBO{45}
\def\Yang{46}
\def\AS{47}
\def\PS{48}
\def\Schulz{49}
\def\JSuz{50}
\def\RB{51}
\def\Kl{52}
\def\SuzGtwo{53}
\def\KS{54}
\def\KOS{55}
\def\BazR{56}
\def\KP{57}
\def\Martinss{58}
\def\ZB{59}
\def\Rava{60}
\def\KPHX{61}

%
%
\centerline{\bf  From Fusion Hierarchy to Excited State TBA }\pn
\vskip 1.2cm
\centerline{
 G. J\"uttner$^{a,1}$
 \footnote{\null}
 {$^1$ E-mail gj@thp.uni-koeln.de}, 
 A. Kl\"umper$^{a,2}$ 
\footnote{\null}
 {$^2$ E-mail kluemper@thp.uni-koeln.de}
and
 J. Suzuki$^{b,3}$
\footnote{\null}
 {$^3$ E-mail suz@hep1.c.u-tokyo.ac.jp}
}\pn
\vskip 1.2cm
\centerline{\sl \phantom{}$^a$ Universit\"at zu K\"oln,            
        Institut f\"ur Theoretische Physik}  \pn  
\centerline{\sl Z\"ulpicher Str. 77,  D-50937, Germany}\pn
\centerline{\sl \phantom{}$^b$ Institute of Physics,            
        University of Tokyo at Komaba}  \pn  
\centerline{\sl Komaba 3-8-1, Meguro-ku, Tokyo, Japan}\pn
\vskip 1.2cm
\noindent Abstract: 
Functional relations among
the fusion hierarchy of quantum transfer matrices give
a novel derivation of the TBA equations, 
namely without string hypothesis.
 This is demonstrated for two important
models of 1D highly correlated electron systems, 
the supersymmetric $t-J$ model and the supersymmetric extended
Hubbard model.
As a consequence, ``the excited state TBA" equations,
which characterize correlation lengths, are
explicitly derived for the $t-J$ model.
To the authors' knowledge, this is the first explicit derivation of
excited state TBA equations for 1D lattice electron systems.
\vfill\eject
%
%
\beginsection{1. Introduction }

The formulation and application of the string hypothesis has 
been a story of success as well as a longstanding mystery 
in the studies of exactly solvable models
[\YY,\Gaud,\HeisenTak,\TakSuz].
For finite systems, the deviations from this hypothesis
have been discussed [\Vla,\KluZitt,\EKSstring,\JuDo] 
and prove essential for a proper description
of critical properties [\spinkb,\spinkbt,\AlczMart].
On the other hand, for finite temperatures
we expect the hypothesis to give the
correct free energy in the thermodynamic limit [\HeisenTak,\TakSuz].

An alternative approach to thermodynamics which
is free of the string hypothesis has been proposed 
[\MSuz,\InSuz,\Koma,\SAW,\SNW,\TakQT,\Klu,\KZeit,\DdV,\Mizu].
There the essential idea is
to deal with 2D classical counterparts.
A novel object, ``the quantum transfer matrix" plays
a fundamental role.
Remarkably, the problem of summing up
all states reduces to the evaluation of only the largest
eigenvalue of the quantum transfer matrix. Of course,
the remaining task is the diagonalization of the quantum transfer matrix.
In order to achieve this, a more sophisticated approach has been proposed
recently by incorporating the classical integrability structure 
in two dimensions [\Klu,\KZeit,\KWZ,\JK,\JKSone,\JKStwo].
The difficulty still remains in the fact that
the resultant system is virtually of finite size with
interactions depending on the size.

Locating the Bethe ansatz roots for finite system sizes is
another difficult task, which we avoid here by one of the
following two alternative strategies.
We encode the information of BAE roots for finite size systems
by 

\item{(1)} a finite number of coupled non-linear integral equations
for a finite number of unknown functions derived within a direct
Bethe ansatz for the diagonalization problem, or
\item{(2)} an infinite number of coupled non-linear integral equations
for an infinite number of unknown functions obtained within the
fusion approach to the diagonalization problem.

\noindent
Note that both transformations are exact.
However, approach (1) has
an evident advantage in numerical investigations.
Actually several thermodynamical quantities have been
explicitly obtained for extensive ranges of temperature
and chemical potentials for 1D correlated electron
systems within this scheme [\JK,\JKSone,\JKStwo].
The latter approach (2) would be less important in 
actual calculations. Nevertheless, it is interesting in its 
connection with the traditional string approach.
The infinite number of unknown functions introduced there,
physically corresponds to the fused (``higher spin") transfer matrices.
These transfer matrices satisfy algebraic functional relations.
For the simplest case, these relations are shown to have
a deep mathematical origin, an exact sequence of 
Yangian modules [\Cherednik,\CPo,\CPt,\KNS].

In this paper we will show that
{\it  these functional relations can be transformed into
      non-linear integral equations, which  are identical to
      TBA equations}.
Note that our approach is completely independent of the string hypothesis.
It thus gives an independent derivation and 
further support for the validity of the
resultant TBA equations.
For the physical applications we need only the
eigenvalue of the fundamental (``un-fused'') quantum transfer
matrix.
Thus the introduction of the  whole commuting
family of transfer matrices seems to be redundant.
The point is that the exact relation among them
makes the evaluation of the fundamental one possible.
This fact was first realized in [\Klu] for the RSOS chains.
In this paper, we will apply this scheme to two 
1D highly correlated electron models to demonstrate that
the relation of the fusion analysis of the quantum transfer matrix and
the TBA is not accidental but universal. \pn

The novel formalism, utilizing the functional relations,
not only reproduces the results obtained from the string hypothesis,
but goes well beyond.
TBA equations usually deal with the equilibrium free energy, 
i.e.\ we can only deal with the ``largest eigenvalue
sector" in the sense of the quantum transfer matrix.
The functional relations are, on the other hand, 
valid for arbitrary eigenvalue sectors.
The latter, therefore offers a wider possibility in calculating
various physical quantities.
Actually, we need to evaluate the subleading eigenvalues
for obtaining correlation lengths.
Recently, some attempts have been made in the context of deformed
conformal field theories, to modify TBA equations so as to
deal with excited states [\Martins,\Fendley,\KN,\Klu,\KNSCh,\TateoDorey,\BLZ].
We will show how the quantum transfer matrix formalism 
naturally leads to excited state TBA equations.
To our knowledge, this is the first explicit derivation of
the excited state TBA equations for 1D lattice electron systems.

This paper is organized as follows.
In the next section, we will give a brief review of
the quantum transfer matrix approach to one-dimensional
quantum systems at finite temperatures. 
In section 3, the functional relations among the relevant transfer
matrices are given. 
The corresponding relations for the
ordinary row-to-row transfer matrices were obtained recently [\Tsuboi].
Here, we will summarize the necessary modifications for the quantum transfer 
matrices.
Section 4 and 5 are devoted to the thermodynamics of 
the integrable $t-J$ model and the supersymmetric extended Hubbard
model, respectively.
We will show explicitly how the functional relations of
section 3 can be transformed in the case of the largest eigenvalue
into the TBA equations of [\Schlottmann,\EK].
To this end, we adopt assumptions about the analyticity of auxiliary
function, which are checked numerically for the case of finite 
Trotter numbers.
In section 6, we will discuss ``excitations at finite temperatures'',
i.e. the correlation lengths of static two point functions.
Assisted by numerical investigations,
the excited state TBA equations are easily derived there.
Section 7 is devoted to the summary and discussion.
%
%

\beginsection{2. The quantum transfer matrix formalism}

In the following we will mainly focus on
two specific models of 1D highly correlated electron
systems: the supersymmetric $t-J$ model [\SuthTJ,\Schlottmann,\BB,\BBO] 
and the supersymmetric extended Hubbard (SEH) model [\EKS].
They describe spin-1/2 electron systems on the lattice.
The prototype of such systems is the celebrated
Hubbard model.
The supersymmetric $t-J$ model may be viewed
as the large Coulomb repulsion limit of a generalized Hubbard model.
The SEH model on the other hand, generalizes the
Hubbard model by additional bond-charge
interactions etc. It shares the same interesting physical
symmetry with the standard Hubbard model resulting in the
eta pairing [\Yang,\AS].
Mathematically the supersymmetric $t-J$ model and SEH model
are in the category of
models based on  Lie superalgebras $gl(r|s)$ with
$r=2$, $s=1$ for
the $t-J$ model and $r=s=2$ for the SEH model.
As remarked in [\JSuz,\RB,\KWZ], 
their classical counterparts are special cases
of the Perk-Schulz [\PS,\Schulz] model with rational
vertex weights ${\cal R}_{\alpha\beta}^{\mu\nu}$
$$\eqalignno{
    {\cal R}_{\alpha\alpha}^{\alpha\alpha}(v)&= 
    1+\epsilon_\alpha v,&\cr
    {\cal R}_{\alpha\alpha}^{\mu\mu}(v)&=
    \epsilon_{\alpha}\epsilon_{\mu}v,&{\R}\cr
    {\cal R}_{\mu\alpha}^{\alpha\mu}(v)&= 
    1,&
}
$$
which satisfy the Yang-Baxter equation. 
As a consequence the row-to-row
transfer matrix ${\cal T}(v)$ constitutes a commuting family with 
respect to the
spectral variable $v$. $\epsilon_\alpha=\pm 1$ are discrete parameters 
determining the `grading' of the system.
The Hamiltonian of the associated quantum system
is given by the logarithmic derivative
$$
  {\cal H}={{d}\over{dv}}\ln {\cal T}(v){|_{v=0}}.\eqno{\Ham}
$$
The supersymmetric $t-J$ Hamiltonian is obtained for 
($\epsilon_1, \epsilon_2, \epsilon_3$) = ($+, -, +$). Likewise the 
grading $(\epsilon_1,\epsilon_2,
\epsilon_3,\epsilon_4 )=(+,-,-,+)$ leads to the SEH model.

Next we introduce Boltzmann weights $\overline{{\cal R}}$ and 
$\widetilde{{\cal R}}$
related to ${{\cal R}}$ by anti-clockwise and clockwise
$90^0$ rotations
$$
  \overline{{\cal R}}_{\alpha\beta}^{\mu\nu}(v)=
  {{\cal R}}^{\alpha\beta}_{\nu\mu}(v), \qquad
  \widetilde{{\cal R}}_{\alpha\beta}^{\mu\nu}(v)=
  {{\cal R}}^{\beta\alpha}_{\mu\nu}(-v).\eqno{\Rmod}
$$
According to \R,\Ham\ we find
$$
  {\cal T}(v)={\cal T}_R\, \e^{v{\cal H}+{\cal O}(v^2)},\qquad
  {\overline{\cal T}}(v)={\cal T}_L\, \e^{v{\cal H}+{\cal O}(v^2)},
\eqno{\trotter}
$$
where ${\cal T}_{R,L}$ are right and left-shift operators. 
Therefore the  partition function  of the quantum system is given by
[\SAW,\Klu,\DdV]
$$
  Z=\lim_{N\to\infty}{\rm Tr}
  \big[{\cal T}(u)\overline{{\cal T}}(u)\big]^{N/2},\quad u=-\beta/N.
\eqno{\staggered}
$$
The right-hand side of this equation may be viewed as the partition 
function of a staggered Perk-Schulz model consisting of alternating rows of 
${{\cal R}}(u)$ and $\overline{{\cal R}}(u)$ operators. For the 
following analytic calculations the 
column-to-column transfer matrix ${\cal T}^{QTM}$ (quantum transfer matrix) 
is best adapted. For quite general systems it can be shown that 
${\cal T}^{QTM}$
possesses a gap between the largest and next-largest eigenvalues
$\Lambda_{\rm max}$ and $\Lambda_{\rm next}$ persisting in the limit 
$N\to\infty$. 
Therefore, the problem of computing the free energy $f$ per site of the
quantum system at finite temperature is reduced to just the
evaluation of the largest
eigenvalue $\Lambda_{\rm max}$
$$
f=-k_B T \ln\Lambda_{\rm max},\qquad (N\to\infty).\eqno{\free}
$$
If also the gap can be calculated (as shown below) the correlation
length at finite temperature can be derived
$$
\xi=\left(\ln{\Lambda_{\rm max}\over\Lambda_{\rm next}}\right)^{-1},
\qquad (N\to\infty).\eqno{\corrlength}
$$

The quantum transfer matrix is given by 
an alternating product of ${{\cal R}}(u)$ and $\widetilde{{\cal R}}(-u)$
operators. The important observation [\Klu,\Kl] is that
$\widetilde{{\cal R}}(v)$ and ${{\cal R}}(v)$ share
the same intertwiner as can be proven most easily in a graphical
way, see Fig.1. 
Therefore 
$$
  {\cal T}^{\rm QTM}(u,v)=\prod_{i=1}^{N/2\ \otimes}
  {\cal R}(v+u)\otimes\widetilde{{\cal R}}(v-u),\eqno{\embedding}
$$
represents a family of commuting matrices comprising the `physical' QTM
at $v=0$.
This intertwining property implies that the Yangian algebra, 
in the realization $RLL = LLR$, is identical for both monodromy matrices.
Consequently, we should have the same functional
relations among the transfer matrices in spite of
apparent differences between the explicit eigenvalues.
Of course this argument needs further elaboration.
However, we leave it as an interesting future 
problem and assume the validity of the above argument
in the following.
%
%
\beginsection{3. The fusion hierarchy}

First, we like to sketch the strategy for the diagonalization of the
quantum transfer matrix for the
Perk-Schulz model spectral parameters $u, iv$
(where we have introduced the factor $i$ for later convenience). 
In the following analysis we will be dealing with transfer matrices
formulated for fused Boltzmann weights. These matrices are denoted
by $T^{(a)}_m(u,v)$ where $T^{(1)}_1(u,v)={\cal T}^{\rm QTM}(u,iv)$.
Of course $T^{(1)}_1(u,0)$ is the only quantity we are 
interested in. For calculating the eigenvalues of $T^{(1)}_1(u,0)$
it is essential to deal with all $T^{(a)}_m(u,v)$ as our method employs
functional equations dealing with the dependence on $v$ and all other
matrices for any $a, m$ [\Klu]. \par
In order to present our results for the eigenvalue of the
quantum transfer matrix we adopt a compact notation using
the ``Yangian analogue of Young tableaux" [\SuzGtwo,\KS,\KOS].
First consider the simplest case:
the quantum transfer matrix
with $r+s$ dimensional degrees of freedom on both
vertical and horizontal edges.
Let us introduce boxes with a letter $\in \{1, \cdots, r+s \}$.
Each box corresponds to an expression as follows.
$$
\Fsquare(0.5cm, \rm{a})_v
=f_a(u+iv) g_a(u-iv) \epsilon_{a}^{N_{a-1}+N_a}
  {{Q_{a-1}(v-i\epsilon_a)} \over {Q_{a-1}(v)}}
  {{Q_{a}(v+i\epsilon_a)} \over {Q_{a}(v) }} ,
\eqno{\boxo}
$$
where
$$\eqalign{
f_a(x) &= \cases{ (1+\epsilon_1 x)^{N/2},&  for $a=1$, \cr
                    x^{N/2},            &  otherwise, \cr}   \cr  
g_a(x) &= \cases{ (1+\epsilon_{r+s} x)^{N/2},&  for $a=r+s$, \cr
                    x^{N/2},            &  otherwise, \cr}   \cr       
Q_{a}(x) &=\cases{ \prod_{j=1}^{N_a} (x-x_j^{(a)}),& for $a=1, 
\cdots, r+s-1$, \cr
                    1,&                              for $a=0,  r+s$. \cr}
}$$
Note that a box carries the spectral parameter dependence.
In the last equation, $x_j^{(a)} (j=1,\cdots, N_a)$ 
denotes the  Bethe ansatz roots with ``color" $a$.
(We adopt the convention $N_0=N_{r+s}=0$).
As derived by an algebraic Bethe ansatz the 
eigenvalue $ \Lambda^{(1)}_1(u,v)$ of the
quantum transfer matrix is of the form
$$
 \Lambda^{(1)}_1(u,v)= \sum_{a=1}^{r+s} \Fsquare(0.5cm,\rm{a})_v.
\eqno{\eig}
$$

The Bethe ansatz equations are obtained by the
pole free condition for the eigenvalue. The BAEs are therefore given by
$$\eqalignno{
\phantom{}&\hbox{Res}_{v=x_j^{(a)}}
(\Fsquare(0.5cm,\rm{a})_v+ \Frect(0.5cm,0.8cm,\rm{a+1})_v ) =0,  \cr
\Longleftrightarrow&   \phantom{}\cr
\phantom{}& -\epsilon_a^{N_{a-1}+N_a} \epsilon_{a+1}^{N_{a}+N_{a+1}}
{{f_a(u+ix_j^{(a)})g_a(u-ix_j^{(a)}) }
  \over {f_{a+1}(u+ix_j^{(a)})g_{a+1}(u-ix_j^{(a)})}}
    =  \cr
&{{Q_{a-1}(x_j^{(a)}) 
  Q_{a}(x_j^{(a)}-i\epsilon_{a+1}) Q_{a+1}(x_j^{(a)}+i\epsilon_{a+1}) }
 \over
  {Q_{a-1}(x_j^{(a)}-i\epsilon_{a}) 
   Q_{a}(x_j^{(a)}+i\epsilon_{a}) Q_{a+1}(x_j^{(a)}) }}  &\baeo
}$$
Starting from $\Lambda^{(1)}_1 $ as in [\KS], one can generate
a set of analytic functions under the BAE.
The resultant  functions  are expected to be  eigenvalues of
fusion transfer matrices.
 For the simplest case like $sl_2$ spin chains,
we can explicitly prove this [\BazR,\KP].
We do not attempt to prove this observation for the general case as our 
analytic treatment below does not depend on it. 
In the case of  row-to-row transfer matrices, the ``Bethe-strap procedure"
has been executed for $sl(r|s)$ models [\Tsuboi].
The eigenvalues of fusion transfer matrices are parameterized by a
similar set of tableaux under certain combinatorial rules.
A set of functional relations among them is also found.
As remarked above, the same set of functional relations holds
for the quantum transfer matrices as well as
the combinatorial rules
for the tableaux giving the analytic eigenvalues for the fusion 
transfer matrices.
 If one replaces boxes for
the row-to-row case by the quantum ones in eq \boxo, then the
explicit eigenvalues of  ``fused transfer matrices'' are obtained.
Let us further explain the rules in the present context.
Consider an $a \times m$ rectangular tableaux.
We fix the ``coordinate" so that the upper-leftmost box
is in the position (1,1), the one box to the right is in (1,2)
and the one box lower is in (2,1).
We associate the spectral parameter $v+i(m-a)/2+i(j-k) $
with the box in $(j,k)$.
Denote the letter in the $(j,k)$ box by $\ell_{j,k}$.
We identify such tableaux with the expression
$$
\prod_{j=1}^{a} \prod_{k=1}^{m}
  \Frect(0.5cm, 0.8cm,\ell_{j,k})_{v+i(m-a)/2+i(j-k)}. 
\eqno{\prodbox}
$$ 
To describe the combinatorial rules
we introduce the order among letters,
$1 \prec 2 \prec \cdots \prec r+s$.
These letters are further classified into
two subsets $J_{+}$ and $J_{-}$.
Then the admissibility conditions are [\Tsuboi] 

\item{(1)} $ \ell_{j-1,k} \prec \ell_{j,k} \quad \hbox{if  } 
               \ell_{j,k} \in J_{+}  $
\item{(2)} $ \ell_{j-1,k} \preceq \ell_{j,k} \quad \hbox{if  } 
              \ell_{j,k} \in J_{-}  $
\item{(3)} $\ell_{j,k-1}\preceq \ell_{j,k} \quad \hbox{if  }
              \ell_{j,k} \in J_{+}  $
\item{(4)} $\ell_{j,k-1}\prec \ell_{j,k} \quad \hbox{if  }  
             \ell_{j,k} \in J_{-} $ 

Define $\Lambda_{m}^{(a)}(u,v)$ by
$$
\Lambda_{m}^{(a)}(u,v)=
\sum_{\{\ell_{j,k} \} \hbox{ under rules (1)-(4) }} 
 \prod_{j=1}^{a} \prod_{k=1}^{m}
  \Frect(0.5cm, 0.8cm,\ell_{j,k})_{v+i(m-a)/2+i(j-k)}.
\eqno{\lambox}
$$
$\Lambda_{m}^{(a)}(u,v)$ is expected to be 
the eigenvalue of the quantum transfer matrix
whose trace has been taken over the Yangian module $W^{(a)}_m$.
We have one further rule [\Martinss]:
$$
\Lambda_{m}^{(a)}(u,v)=0 \quad \hbox{if } a > r \hbox{ and } m>s.
\eqno{\condo}
$$ 
The functional relations read
$$
\Lambda_{m}^{(a)}(u,v-{i \over 2}) \Lambda_{m}^{(a)}(u,v+{i \over 2})=
\Lambda_{m}^{(a+1)}(u,v)\Lambda_{m}^{(a-1)}(u,v)+
\Lambda_{m-1}^{(a)}(u,v)\Lambda_{m+1}^{(a)}(u,v).
\eqno{\funco}
$$
with the convention $\Lambda_{0}^{(a)}=\Lambda_{m}^{(0)}=1$.
These relations can be proved using a quantum analogue of
the Jacobi-Trudi formula and Pl{\"u}cker's relation [\KOS,\Tsuboi].
Recently
some attempts have been made in the study of finite-size corrections
in $t-J$ like models employing similar functional relations 
 [\ZB]. There the row-to-row transfer matrix was considered and a 
subset of fusion functions was used.

In the next two sections, we will transform the complete set
of functional equations 
into integral form which is useful for further analytic
treatment. 
We will show explicitly how the 
simple functional relations above are nothing but TBA equations
for the quantum chains related to the Perk-Schulz model.
We will focus on two specific examples of interests, 
the integrable $t-J$ model and the SEH model.
%
%
%
%
\beginsection{4. 
   Fusion based derivation of TBA for the
   supersymmetric $t-J$ model }

For the $t-J$ model, we choose a convenient grading $\epsilon_1=\epsilon_3=1,
\epsilon_2=-1$. Then $J_+=\{1,3\}$, $J_{-}=\{2\}$.
For example, we have the following five tables for $\Lambda^{(1)}_2$
$$
\Fsquare(0.5cm, 1)\Addsquare(0.5cm,1), \quad
\Fsquare(0.5cm, 1)\Addsquare(0.5cm,2), \quad
\Fsquare(0.5cm, 1)\Addsquare(0.5cm,3), \quad
\Fsquare(0.5cm, 2)\Addsquare(0.5cm,3), \quad
\Fsquare(0.5cm, 3)\Addsquare(0.5cm,3),
$$
and four tables for $\Lambda^{(2)}_1$,
$$
\hbox{
   \normalbaselines\m@th\baselineskip0pt\offinterlineskip
   \vbox{ 
      \hbox{$\Fsquare(0.5cm,1)$}\vskip-0.4pt
             \hbox{$\Fsquare(0.5cm,2)$}\vskip-0.4pt
        }
      }
\quad
\hbox{
   \normalbaselines\m@th\baselineskip0pt\offinterlineskip
   \vbox{ 
      \hbox{$\Fsquare(0.5cm,1)$}\vskip-0.4pt
             \hbox{$\Fsquare(0.5cm,3)$}\vskip-0.4pt
        }
      }
\quad
\hbox{
   \normalbaselines\m@th\baselineskip0pt\offinterlineskip
   \vbox{ 
      \hbox{$\Fsquare(0.5cm,2)$}\vskip-0.4pt
             \hbox{$\Fsquare(0.5cm,2)$}\vskip-0.4pt
        }
      }
\quad
\hbox{
   \normalbaselines\m@th\baselineskip0pt\offinterlineskip
   \vbox{ 
      \hbox{$\Fsquare(0.5cm,2)$}\vskip-0.4pt
             \hbox{$\Fsquare(0.5cm,3)$}\vskip-0.4pt
        }
      },
$$
and so on.
From the analytic point of view, it is better to divide the
$\Lambda^{(a)}_m(u,v)$ as defined in \lambox \  by a common factor, such
that the resultant expression is a  polynomial  of degree
$N$ in $v$ irrespective of the values of $a$ and $m$.
For that purpose we prepare the normalization functions,
$$\eqalignno{
h^{(1)}_m(x) =&\prod_{j=1}^{m-1} \phi_-(x+(j-{{(m+1)}\over2})i)
                           \phi_+(x-(j-{{(m+1)}\over2})i) ,
                                       &\hdef        \cr
n^{(a)}_m(x) =
         &\prod_{j=1}^a h^{(1)}_m(x+(j-{{a+1}\over 2})i)\times&\cr
         & \prod_{j=1}^{a-1}\phi_-(x+{{(m+a-2j)}\over 2}i)  
                           \phi_+(x-{{(m+a-2j)}\over 2}i) ,\qquad
                                       &\ndef        \cr
}$$
where  $\phi_{\pm}(v)= (v \pm iu)^{N/2}$.
Then the normalized functions $\widetilde{\Lambda^{(a)}_m}$ are given by
$$\eqalignno{
\Lambda^{(a)}_m (u,x) &= n^{(a)}_m(x) \widetilde{\Lambda^{(a)}_m}(x) ,
                      &\lamamtj\cr
\widetilde{\Lambda^{(0)}_m}(x) &= 
               \phi_+(x-{m \over 2}i)  \phi_-(x+{m \over 2}i),
                      &\lam0mtj\cr
\widetilde{\Lambda^{(a)}_0}(x) &= 
               \phi_+(x+{a \over 2}i)  \phi_-(x-{a \over 2}i) .
                     &\lama0tj
}$$
Hereafter  we drop $u$ from $\widetilde{\Lambda^{(a)}_m}$ for
simplicity.  \pn
Now the functional relations read,
$$\eqalignno{
\widetilde{\Lambda^{(1)}_m} (x-i/2) \widetilde{\Lambda^{(1)}_m}(x+i/2) &=
    \widetilde{\Lambda^{(1)}_{m+1}}(x) \widetilde{\Lambda^{(1)}_{m-1}}(x) + 
     \widetilde{\Lambda^{(2)}_m}(x)    \widetilde{\Lambda^{(0)}_m}(x),
    \quad  m=1, \cdots , \infty ,\qquad 
               &\freltjo\cr
\widetilde{\Lambda^{(2)}_1}(x-i/2) \widetilde{\Lambda^{(2)}_1}(x+i/2) &=
    \widetilde{\Lambda^{(2)}_{2}}(x)   \widetilde{\Lambda^{(2)}_{0}}(x) + 
    \widetilde{\Lambda^{(1)}_1} (x) \widetilde{\Lambda^{(3)}_1} (x),
                     &\freltjtw \cr
 \widetilde{\Lambda^{(2)}_m}(x-i/2)  \widetilde{\Lambda^{(2)}_m}(x+i/2) &=
    \widetilde{\Lambda^{(2)}_{m+1}}(x)  \widetilde{\Lambda^{(2)}_{m-1}}(x),
    \qquad  m=2, \cdots , \infty,
              &\freltjth \cr
 \widetilde{\Lambda^{(a)}_1}(x-i/2)  \widetilde{\Lambda^{(a)}_1}(x+i/2) &=
     \widetilde{\Lambda^{(a+1)}_1} (x)  \widetilde{\Lambda^{(a-1)}_1} (x),
    \qquad  a=3, \cdots , \infty.
                &\freltjfo \cr
}$$
In the above,   not all $\widetilde{\Lambda^{(a)}_m}(x)$  are
independent; we have the identity
$ \widetilde{\Lambda^{(a+1)}_1}(x)=\widetilde{\Lambda^{(2)}_a}(x)$,
which can be easily proved by explicit forms.
Now we apply transformations generalizing the
ones in  [\Klu,\KP,\KNS]
$$\eqalignno{
Y^{(1)}_m(x) &= 
 {{\widetilde{\Lambda^{(1)}_{m+1}}(x) \widetilde{\Lambda^{(1)}_{m-1}}(x)} 
   \over
  {\widetilde{\Lambda^{(2)}_{m}}(x) \widetilde{\Lambda^{(0)}_{m}}(x)}  },
  \qquad \hbox{for  } m \ge 1,
                &\ytranstjo\cr
Y^{(a)}_1(x) &={ {\widetilde{\Lambda^{(a)}_0}(x)}\over
                     { \widetilde{\Lambda^{(a-1)}_1}(x)  }},
  \qquad \hbox{for  } a \ge 2.
               &\ytranstjtw. 
}$$
We call the functional relations in terms of the 
functions $Y^{(a)}_m(x)$ simply the ``Y-system'' [\AlZ,\KM,\Rava]. 
The explicit Y-system  is now given by
$$\eqalignno{
Y^{(1)}_1(x-{i\over 2}) Y^{(1)}_1(x+{i\over 2}) &=
     {{(1+Y^{(1)}_2(x))}\over {(1+(Y^{(2)}_1(x))^{-1})   }  },
            &\ysystjo\cr
Y^{(1)}_m(x-{i\over 2}) Y^{(1)}_m(x+{i\over 2}) &=
    (1+Y^{(1)}_{m+1}(x)) (1+Y^{(1)}_{m-1}(x)),
     \qquad \hbox{for  } m \ge 2,
            &\ysystjtw\cr
Y^{(2)}_1(x-{i\over 2}) Y^{(2)}_1(x+{i\over 2}) &=
  {{\phi_+(x+i/2)\phi_-(x-i/2)} \over {\phi_-(x+i/2)\phi_+(x-i/2)}}
  {{  Y^{(3)}_1(x)}\over{(1+  Y^{(1)}_1(x)  )  }},
            &\ysystjth\cr
Y^{(3)}_1(x-{i\over 2}) Y^{(3)}_1(x+{i\over 2}) &=
  {{  Y^{(4)}_1(x)}\over{(1+  (Y^{(2)}_1(x)  )^{-1}) }},
           &\ysystjfo\cr
Y^{(a)}_1(x-{i\over 2}) Y^{(a)}_1(x+{i\over 2}) &=
Y^{(a+1)}_1(x) Y^{(a-1)}_1(x),    \qquad \hbox{for  } a \ge 4.
          &\ysystjfi\cr
}$$
At this stage 
we note some numerical results concerning the analytic properties of
$\widetilde{\Lambda^{(1)}_m}(x)$. 
They are analytic due to the BAE, and have $N$ zeros.
Through finite $N$ studies, we find their characteristic
behavior.
For the eigenstate giving the 
largest eigenvalue  for  $\widetilde{\Lambda^{(1)}_1}(x)$,
the zeros are located on
curved lines with imaginary parts close to
$\pm (m+1)/2$.
Therefore all $Y^{(a)}_m(x)$ except for $a=m=1$ are Analytic
Nonzero and have Constant asymptotics(ANZC) in the strip
$\Im x \in \{-1/2, 1/2 \}$.
Similarly we have numerically checked that $1+Y^{(a)}_m(x)$ has
the ANZC property in this strip. For the case of $(1+Y^{(1)}_2)$ see
Fig.~2.
To cancel  zeros and holes  for $Y^{(1)}_1(x)$, we
define
$$
 \widetilde{Y^{(a)}_m(x)}= 
\cases{ ( \tanh {\pi\over 2}(x+i({1\over 2}+u))
          \tanh {\pi\over 2}(x-i({1\over 2}+u)) )^{-N/2}\, Y^{(a)}_m(x),
                                 &  \hbox{for  } a=m=1, \cr
               Y^{(a)}_m(x),&     \hbox{otherwise}.   \cr
        }
\eqno{\ytilde}
$$
Now the left hand sides of eqs. \ysys \  are invariant if we replace
$Y^{(a)}_m(x)$  by  $\widetilde{Y^{(a)}_m(x)}$.
After this replacement, both sides of eqs. \ysys\ are ANZC functions
in a narrow strip including the real axis. 
Therefore the Fourier transformation of logarithms of
both sides
can be simplified due to Cauchy's theorem.
This results in a set of linear equations in Fourier space which can
be solved easily. The reverse
Fourier transform leads to the following
infinitely many coupled non-linear integral equations.
$$\eqalignno{
\log Y^{(1)}_1 (x) &= e_1(u,x) + G_0*\log(1+Y^{(1)}_2)(x)
                                 -G_0*\log(1+{1\over{Y^{(2)}_1}})(x),
                    &\ninttjo\cr
\log Y^{(1)}_m (x) &=
  G_0*\log(1+Y^{(1)}_{m-1})(x)+G_0*\log(1+Y^{(1)}_{m+1})(x), \quad
       \hbox{ for  } m\ge 2,
                   &\ninttjtw\cr
\log Y^{(2)}_1 (x) &=
     e_2(u,x) + G_0*\log Y^{(3)}_1(x)
                    -G_0*\log(1+Y^{(1)}_1)(x),
                  &\ninttjth \cr
\log Y^{(3)}_1 (x) &=
   G_0*\log Y^{(4)}_1(x)
                    -G_0*\log(1+{1\over{Y^{(2)}_1}}),
                      &\ninttjfo\cr
\log Y^{(a)}_1 (x) &=
   G_0*\log Y^{(a-1)}_1(x)+ G_0*\log Y^{(a+1)}_1(x),
   \quad \hbox{ for } a \ge 4.
                      &\ninttjfi
}$$
Here $A*B(x)$ means convolution $A*B= \int A(x-y)B(y) dy$, and
$$
\eqalign{
e_1(u,x) &= \log \Bigl(  \tanh{\pi \over 2}(x+i({1\over 2}+u))
                          \tanh{\pi \over 2}(x-i({1\over 2}+u))
                      \Bigr )^{N/2}, \cr
e_2(u,x) &=  N \int {{e^{-|k|/2-ikx} \sinh ku}  \over {2k \cosh k/2}}  
\hbox{dk}, \cr
G_0(x) &= {1 \over {2\pi}} \int  {{e^{-ikx}}\over{2 \cosh k/2}}
         = {1\over{2 \cosh \pi x}}.
}$$
Note that $\log Y^{(3)}_1 (x)$ can be solved in term of
$\log(1+{1\over{Y^{(2)}_1}})$ from the last two relations in the above.
Resubstitution of the result into the third relation above leads to
$$
\log Y^{(2)}_1 (x) =
     e_2(u,x) - G_1*\log(1+{1\over{Y^{(2)}_1}})(x)
                    -G_0*\log(1+Y^{(1)}_1)(x),
 \eqno{\ninttjthd}
$$
where $G_1(x) = {1 \over {2\pi}} \int dk {{e^{-ikx-|k|/2}}\over{2 \cosh k/2}}$.
\pn
Therefore we have the resultant Y-system including
$Y^{(1)}_m, (m=1,\cdots, \infty)$ and $Y^{(2)}_1$.
The quantity of our interest, the free energy, is  given in terms of
$Y^{(a)}_m$ as,
$$
-\beta f =
 \lim_{ N \rightarrow \infty} \log \Lambda^{(1)}_1( u=-{\beta \over N},0)
=\lim_{ N \rightarrow \infty} 
   \log \phi_+(i) \phi_-(-i) - \log Y^{(2)}_1( u=-{\beta \over N},0)
\eqno{\freetj}
$$
Now the limit $N \rightarrow \infty$ is taken
easily by analytic means.
To write down the final result, we make slight changes of notations
$$\eqalignno{
Y^{(1)}_1(x) &= \exp (-\epsilon(x)/T),   &\chgo \cr
Y^{(1)}_m(x) &= \exp (\phi_{m-1}(x)/T),   \qquad m \ge 2,   &\chgtw \cr
Y^{(2)}_1(x) &= \exp (-\psi(x)/T).    & \chgth\cr
}
$$
Then we have
$$
\eqalignno{
\epsilon &= 2\pi G_0 +T G_0 *\log {{(1+e^{\psi/T})}\over {(1+e^{\phi_1/T})}},
               &\scho\cr
\psi  &=2\pi G_1 +T G_0 * \log(1+e^{-\epsilon/T})+
             T G_1*\log(1+e^{\psi/T}),
               &\schtw\cr
\phi_1&=TG_0 *\log[(1+e^{-\epsilon/T})(1+e^{\phi_2/T})],
               &\schth\cr
\phi_m&=TG_0 *\log[(1+e^{\phi_{m-1}/T})(1+e^{\phi_{m+1}/T})],
              &\schfo\cr
f&= 1-\psi(0).
}$$
They are nothing but the TBA equations obtained in [\Schlottmann]
for $x_s=1, A'=0$.  The  shift by unity in the free energy is
due to a trivial shift in the definition of the Hamiltonian.
Thus {\it we have succeeded in deriving these equations independent of
          the string hypothesis}.
In the next section, we further apply the same strategy to
another model in 1D correlated electron systems to
convince ourselves of the universality of the result.

\beginsection{5. TBA for the SEH model }

For this model, we choose the grading $\{\epsilon_1,\epsilon_2,
\epsilon_3,\epsilon_4 \}=\{+,-,-,+\}$.
Then $J_+=\{1, 4\}$, $J_-=\{2,3\}$.
Due to these choices, we have non-vanishing $\Lambda$'s:
$\Lambda^{(1)}_m$, $\Lambda^{(2)}_m$, $\Lambda^{(a)}_1$, 
$\Lambda^{(a)}_2$, ($m=0,\cdots, \infty$,  $a=0,\cdots, \infty$).
While the derivations of TBA based on the string hypothesis
for the $t-J$ and the SEH seem to be different,
we find them quite similar by adopting the present
strategy. 
We thus sketch the relevant definitions and relations
for the SEH model, without going into details.
As for the $t-J$ model,
we re-normalize $\Lambda^{(a)}_m(u,v)$ by
the same $n_m^{(a)}(u,v)$  in \ndef.
The  renormalized functional relations are given by
$$\eqalignno{
\widetilde{\Lambda^{(1)}_m} (x-i/2) \widetilde{\Lambda^{(1)}_m}(x+i/2) &=
    \widetilde{\Lambda^{(1)}_{m+1}}(x) \widetilde{\Lambda^{(1)}_{m-1}}(x) + 
     \widetilde{\Lambda^{(2)}_m}(x)    \widetilde{\Lambda^{(0)}_m}(x),
    \  m=1, \cdots , \infty,     \qquad
          & \frelseho \cr
\widetilde{\Lambda^{(a)}_1}(x-i/2) \widetilde{\Lambda^{(a)}_1}(x+i/2) &=
    \widetilde{\Lambda^{(a)}_{2}}(x)   \widetilde{\Lambda^{(a)}_{0}}(x) + 
    \widetilde{\Lambda^{(a-1)}_1} (x) \widetilde{\Lambda^{(a+1)}_1} (x),
    \qquad a=2,\cdots, \infty, 
         & \frelsehtw \cr
\widetilde{\Lambda^{(2)}_2}(x-i/2) \widetilde{\Lambda^{(2)}_2}(x+i/2) &=
    \widetilde{\Lambda^{(2)}_{3}}(x)   \widetilde{\Lambda^{(2)}_{1}}(x) + 
    \widetilde{\Lambda^{(1)}_2} (x) \widetilde{\Lambda^{(3)}_2} (x), 
         & \frelsehth \cr
 \widetilde{\Lambda^{(2)}_m}(x-i/2)  \widetilde{\Lambda^{(2)}_m}(x+i/2) &=
    \widetilde{\Lambda^{(2)}_{m+1}}(x)  \widetilde{\Lambda^{(2)}_{m-1}}(x),
    \qquad  m=3, \cdots , \infty,
         & \frelsehfo \cr
 \widetilde{\Lambda^{(a)}_2}(x-i/2)  \widetilde{\Lambda^{(a)}_2}(x+i/2) &=
     \widetilde{\Lambda^{(a+1)}_2} (x)  \widetilde{\Lambda^{(a-1)}_2} (x),
    \qquad  a=3, \cdots , \infty.
         & \frelsehfi \cr
}$$
In addition we have the equalities,
$$
  \widetilde{\Lambda^{(2)}_{m}}(x) =  \widetilde{\Lambda^{(m)}_{2}}(x),
\quad m\ge 2.
\eqno{\eqalityseh}
$$
We introduce the $Y$ functions as follows
$$\eqalignno{
Y^{(1)}_m(x) &=
 {{\widetilde{\Lambda^{(1)}_{m+1}}(x) \widetilde{\Lambda^{(1)}_{m-1}}(x)} 
   \over
  {\widetilde{\Lambda^{(2)}_{m}}(x) \widetilde{\Lambda^{(0)}_{m}}(x)}  },
  \qquad \hbox{for  } m \ge 1,
       &\ytranseho\cr
Y^{(2)}_m(x) &= 
 {{\widetilde{\Lambda^{(2)}_{m+1}}(x) \widetilde{\Lambda^{(2)}_{m-1}}(x)} 
   \over
  {\widetilde{\Lambda^{(3)}_{m}}(x) \widetilde{\Lambda^{(1)}_{m}}(x)}  },
  \qquad \hbox{for  } m =1,2,
        &\ytransehtw\cr
Y^{(a)}_1(x) &= 
 {{\widetilde{\Lambda^{(a)}_{2}}(x) \widetilde{\Lambda^{(a)}_{0}}(x)} 
   \over
  {\widetilde{\Lambda^{(a+1)}_{1}}(x) \widetilde{\Lambda^{(a-1)}_{1}}(x)}  },
  \qquad \hbox{for  } a\ge 3.
        &\ytransehth\cr
}$$
Then the following Y-system holds,
$$\eqalignno{
Y^{(1)}_1(x-{i\over 2}) Y^{(1)}_1(x+{i\over 2}) &=
     {{(1+Y^{(1)}_2(x))}\over {(1+(Y^{(2)}_1(x))^{-1})   }  },
   &\ysysseho\cr
Y^{(1)}_2(x-{i\over 2}) Y^{(1)}_2(x+{i\over 2}) &=
     {{(1+Y^{(1)}_3(x))(1+Y^{(1)}_1(x))}\over {(1+(Y^{(2)}_2(x))^{-1})   }  },
  &\ysyssehtw \cr
Y^{(1)}_m(x-{i\over 2}) Y^{(1)}_m(x+{i\over 2}) &=
    (1+Y^{(1)}_{m+1}(x)) (1+Y^{(1)}_{m-1}(x)),
     \qquad \hbox{for  } m \ge 3,
     &\ysyssehth\cr
Y^{(2)}_1(x-{i\over 2}) Y^{(2)}_1(x+{i\over 2}) &=
  {{ (1+ Y^{(2)}_2(x))}\over
      {(1+  (Y^{(3)}_1(x)  )^{-1})(1+  (Y^{(1)}_1(x)  )^{-1}) }},
   &\ysyssehfo \cr
Y^{(a)}_1(x-{i\over 2}) Y^{(a)}_1(x+{i\over 2}) &=
Y^{(a+1)}_1(x) Y^{(a-1)}_1(x),    \qquad \hbox{for  } a \ge 3,
  &\ysyssehfi \cr
Y^{(2)}_2(x-{i\over 2}) Y^{(2)}_2(x+{i\over 2}) &=
  {{\phi_+(x+i)\phi_-(x-i)} \over {\phi_-(x+i)\phi_+(x-i)}}
  {{ (1+ (Y^{(2)}_1(x))^{-1})}\over{(1+  Y^{(1)}_2(x)  )  }}. 
  &\ysyssehsi \cr
}$$
Numerically, both sides of the above equations have the ANZC property in a
narrow strip near the real axis, except for the lhs of the
first equation. 
 As for the $t-J$ model, we cancel zeros and singularities by
multiplying the scalar as in \ytilde.
Then the Y-system is equivalent to the following set of
non-linear integral equations
$$\eqalignno{
\log Y^{(1)}_1 (x) &= e_1(u,x) + G_0*\log(1+Y^{(1)}_2)(x)
                                 -G_0*\log(1+{1\over{Y^{(2)}_1}})(x),
      &\nintseho\cr
\log Y^{(1)}_2 (x) &= G_0*
\log[{{(1+Y^{(1)}_3)(x) (1+Y^{(1)}_1)(x) } \over
       {(1+(Y^{(2)}_1(x))^{-1})}}],     
     &\nintsehtw  \cr
\log Y^{(1)}_m (x) &=
  G_0*\log(1+Y^{(1)}_{m-1})(x)+G_0*(1+Y^{(1)}_{m+1})(x)), \quad
       \hbox{ for  } m\ge 3,
      &\nintsehth  \cr
\log Y^{(a)}_1 (x) &=
   G_0*\log Y^{(a-1)}_1(x)+ G_0*\log Y^{(a+1)}_1(x),
   \quad \hbox{ for } a \ge 3,
      &\nintsehfo\cr
\log Y^{(2)}_2 (x) &=
     e_3(u,x) + G_0*\log (1+(Y^{(2)}_1(x))^{-1})
                    -G_0*\log(1+Y^{(1)}_2)(x),
       &\nintsehfi\cr
\log Y^{(2)}_1 (x) &=
   G_0* \log[{ {(1+Y^{(2)}_2(x))}\over
               { (1+(Y^{(3)}_1(x))^{-1})(1+(Y^{(1)}_1(x))^{-1})}}], \cr
   & = e_2(u,x) +
   G_0* \log[{ {(1+(Y^{(2)}_2(x))^{-1})}\over
               { (1+(Y^{(3)}_1(x))^{-1})(1+Y^{(1)}_1(x))}}],
          &\nintsehsi \cr
 e_3(u,x)&= N \int {{e^{-|k|-ikx} \sinh ku}  \over {2k \cosh k/2}}  \hbox{dk}.
}$$
$\Lambda^{(1)}_1(u,x)$ in terms of $Y^{(a)}_m$ reads,
$$\eqalignno{
\Lambda^{(1)}_1(u,x)&=\log \phi_+(x+i) \phi_-(x-i) +
            \sum_{j=1}^{\infty} K_j* \log( 1+(Y^{(1)}_j)^{-1}),
    & \eigseh \cr
K_j(x) &= {j \over {x^2+(j/2)^2}} .
}$$
Now let $N \rightarrow \infty$ under identifications,
$$\eqalign{
\alpha_n &=Y^{(1)}_n, \quad (n\ge 1);  \qquad (\alpha_0=1),  \cr
\beta_1&=Y^{(2)}_1, \qquad \beta_2=Y^{(2)}_2,  \cr
\gamma_s&=1/Y^{(s+2)}_1.
}$$
Then the resultant TBA equations are nothing but
the ones in [\EK],
after taking some appropriate linear combinations
$$\eqalignno{
\log \alpha_n &=
  + G_0*\log[(1+\alpha_{n+1})(1+\alpha_{n-1})] \cr
 & -\delta_{n,1} (2\pi G_0/T +\log(1+1/\beta_1)
   -\delta_{n,2}\log(1+1/\beta_2),    \quad n\ge 1, \cr
\log \gamma_s &=
  + G_0*\log[(1+\gamma_{s+1})(1+\gamma_{s-1})] 
  +\delta_{s,1} G_0*\log(1+1/\beta_1),  \quad s\ge 1, \cr
\log\beta_1  &= -2\pi G_0/T+
   G_0*\log[{{(1+\beta_1)}\over{(1+\alpha_1)(1+\gamma_1)}} ],   \cr
\log\beta_2  &= -2\pi G_1/T+
   G_0*\log[{{(1+1/\beta_1)}\over{(1+\alpha_2)}}     ],    &\chgseh \cr
f &= 1 -\sum_n K_n * \log(1+1/\alpha_n).
}$$
Note that the present choice of Hamiltonian 
corresponds to
$\mu+U/2=0$.
%
%
\beginsection{6. Excited state TBA }

So far we have shown that the quantum transfer matrix formalism
for the largest eigenvalue of $\Lambda^{(1)}_1$
with adequate information on analyticity
properties
directly relates the functional relations with the TBA equations.
The investigation of other eigenvalues naturally leads to
the excited state TBA equations, which characterize 
correlation lengths at finite temperatures. 
Note that the functional relations are valid  irrespective of
the sectors under consideration.
The only difference to the ``ground state" is the modification
of the analyticity conditions by
additional zeros or singularities entering into
the strips where  auxiliary functions in the ground state are 
strictly analytic and non-zero [\KP,\KPHX].
Let us examine explicitly 
the spin excitation for the solvable $t-J$ model.

For the ground state, we have ${N\over 2}  x^{(1)}_j $ and
 ${N\over 2}  x^{(2)}_j $ Bethe ansatz roots complex conjugate to each other.
This time, we also have such complex conjugate pairs of BAE solutions,
however, the number of solution is $N/2-1$ for each.
%
%
This does not seem to be a serious modification of the ground state,
however it results into considerable differences in the zeros
for $\Lambda^{(a)}_m$.
We find $N-2$ zeros for $\Lambda^{(1)}_m$ on slightly curved lines
close to $\Im x  \sim \pm(m+1)/2 $ as for the ground state.
Note that two zeros are missing on these lines. 
Instead we find (by numerical investigations for vanishing chemical 
potential) the additional structure:
\item{(1)} $\Lambda^{(1)}_m$ has zeros at $x^{(1)}_{m,\pm}$ on the real axis.
The strictly largest eigenvalue is given by two symmetrically distributed
zeros $x^{(1)}_{m,\pm}=\pm x^{(1)}_{m}$.
\item{(2)} $\Lambda^{(a)}_1, a \ge 2$, has zeros 
             at $\pm (x^{(2)}_1 +(a-2)i/2) $ on the imaginary axis. 
We find $x^{(2)}_1$ is pure imaginary and $| x^{(2)}_1|> 1/2 $.\par
\noindent 
As a corollary from the identity 
$ \widetilde{\Lambda^{(2)}_m}= \widetilde{\Lambda^{(m+1)}_1}$ we get
\item{(3)} $\Lambda^{(2)}_m$ has zeros 
             at $\pm (x^{(2)}_1 +(a-1)i/2) $ on the imaginary axis.\par
\noindent 
Note that the behavior of zeros (2) is consistent with the
``trivial'' T-system, 
$$
 \widetilde{\Lambda^{(a)}_1}(x-i/2)  \widetilde{\Lambda^{(a)}_1}(x+i/2) =
     \widetilde{\Lambda^{(a+1)}_1} (x)  \widetilde{\Lambda^{(a-1)}_1} (x),
\,\, a=3, \cdots , \infty.$$
Thus, in terms of the $Y$ functions, we find the following zeros 
and singularities
in the strip $\Im x \in \{-1/2, 1/2 \}$,
$$
\eqalignno{
Y^{(1)}_m(x) &=0,
\qquad\hbox{ at } x=  x^{(1)}_{m-1,\pm}, x^{(1)}_{m+1,\pm} ,\cr
Y^{(2)}_1(x) &= \infty, \qquad \hbox{ at }  x=  x^{(1)}_{1,\pm} .
       &\excyzero
}
$$
In order to have ANZC functions in the strip we must modify the
$\widetilde{Y^{(a)}_m}$ functions of section~4 further to
$$\eqalignno{
\widetilde{Y^{(1)}_m} &\rightarrow \widetilde{Y^{(1)}_m} 
     \prod_{\sigma=\pm}\tanh {\pi \over 2}(x-x^{(1)}_{m+1,\sigma})
     \tanh {\pi \over 2}(x-x^{(1)}_{m-1,\sigma}), \cr
\widetilde{Y^{(2)}_1}  &\rightarrow \widetilde{Y^{(2)}_1} 
      / \prod_{\sigma=\pm}\tanh{\pi \over 2}(x-x^{(1)}_{1,\sigma}).
&\excrenorm
}$$
The resultant $\widetilde{Y^{(a)}_m}$ are now ANZC functions
in the strip.
Repeating the same argument as in section~4, we thus arrive at
$$\eqalignno{
\log Y^{(1)}_1 (u,x) &= e_1(u,x) +
     G_0*\log{{(1+Y^{(1)}_2)}\over{(1+1/Y^{(2)}_1)}} (x)+
    \sum_{\sigma=\pm}\log (\tanh {\pi \over 2}(x-x^{(1)}_{2,\sigma}) ),\qquad
            &\excTBAo\cr
\log Y^{(1)}_m (u,x) &=
  G_0*\log(1+Y^{(1)}_{m-1})(1+Y^{(1)}_{m+1}))(x)  \cr
 & + \sum_{\sigma=\pm}\log (\tanh {\pi \over 2}(x-x^{(1)}_{m+1,\sigma} ) 
             \tanh {\pi \over 2}(x-x^{(1)}_{m-1,\sigma} ) ), \quad
\hbox{        for  } m\ge 2,\qquad
           &\excTBAtw\cr
\log Y^{(2)}_1 (u,x) &=
     e_2(u,x) - G_1*\log(1+{1\over{Y^{(2)}_1}})(x)
              - G_0*\log(1+Y^{(1)}_1)(x)       \cr
    \phantom{  }&-
\sum_{\sigma=\pm}\log(\tanh{\pi \over 2}(x-x^{(1)}_{1,\sigma})),
          &\excTBAth\cr
}$$
and the $x^{(1)}_{m,\sigma} $ satisfy,
$$
 Y^{(1)}_m(u, x=x_{m,\sigma} \pm i/2) = -1.
\eqno{\condxm}
$$
In the new integral equations the convolutions are still evaluated
with integral contours along the real axis.
In this way, we have derived quite naturally excited state TBA equations for
spin excitations. We want to stress once more the meaning of these excitations.
The corresponding eigenvalue determines via \corrlength\  the decay of
static correlation functions, it is not related to dynamical correlations
or ``excited energies at finite temperature'' [\YY].
For other types of excitations, one needs to identify the patterns and
locations of zeros and singularities of the auxiliary functions.
This is  possible for small Trotter numbers,
as we know explicitly the spectrum from ``brute force''
numerical calculations.
We expect the so obtained patterns would be valid also in
the limit $ N \rightarrow \infty$.
Lastly, we want to comment on our derivation of the TBA equations on the
basis of ANZC properties. It appears that any solution to the integral
equations is a solution of the functional equations and thereby yielding
an eigenvalue to the quantum transfer matrix. The question 
remains whether the obtained eigenvalues are the largest ones. This however,
is a byproduct of this section. Any time we modify the ANZC properties
by allowing zeros to enter the physical strips the eigenvalue is lowered.
This is the analytic justification of the previously employed analytic
properties of the largest eigenvalue.

%
\beginsection{5. Summary and Discussion }

In this paper, we have shown the equivalence of the fusion relations
among the quantum transfer matrices and the TBA equations
under appropriate assumptions about analyticities of auxiliary functions.
This has been demonstrated by adopting two examples of 1D highly correlated
electron systems. \par
In our approach the derivation of the TBA equations is based essentially on
ANZC properties of the auxiliary functions satisfying the functional relations.
In some sense, the string hypothesis is now replaced by 
ANZC assumptions.
We like, however, to point out the following. First of all, the analyticity
properties are easier to control. We have checked numerically the ANZC 
properties for various configurations and for several Trotter numbers up
to $N=32$. Second,
in contrast to deviations from the string hypothesis for
the finite system, we do not observe  numerically any violation of 
the latter assumption for finite
Trotter numbers.  \par 
Another advantage in our approach is the straightforward 
derivation of ``excited state TBA''. 
The additional chemical potential terms in TBA equations 
in the CFT limit of
previous studies [\Martins,\Fendley,\KM,\KN] is now
replaced by the concrete objects, the contribution
by the additional zeros $x^{(1)}_m$ of the fusion transfer matrices.
\par
For the actual calculation of physical quantities such as 
the specific heat or correlation lengths,
we must deal with, in principle, an infinite number of
unknown functions and parameters $x^{(1)}_m$.
Let us sketch the situation in the case of the $t-J$ model.
In order to solve numerically the TBA equations \ninttjo - \ninttjfi \ 
we truncate the infinite set of integral equations \ninttjtw \  
to a finite one by 
approximating $\log Y^{(1)}_m(x)$, for $m \gg 1$ by its asymptotic value. 
 We thus introduce the quantity 
$\log(Y^{(1)}_m(0)/Y^{(1)}_m(\infty))$ as
 an estimate for the systematic error. 
To evaluate this, we use the explicit solutions
to BAE for finite Trotter numbers $N$ together with 
functional relations, thereby allowing 
the evaluation of these quantities with high  numerical precision.
In the limit $m\gg 1$ we find the following asymptotic behavior
$$ \log(Y^{(1)}_m(0)/Y^{(1)}_m(\infty)) \sim C/m^2. $$
For the example $N=12$, $\beta=1/T=3.5$ (and vanishing
external fields), we have $C\approx 6$. 
Thus we need more than 2500 auxiliary functions $Y^{(1)}_m$ 
to obtain a numerical error less than $10^{-6}$.
This already may be a sign of difficulties in 
obtaining good numerical accuracy for the traditional TBA analysis.\par
Fortunately, as we already remarked in the introduction,
we have an alternative approach where
only a finite number of auxiliary functions is
necessary [\KZeit,\KWZ,\JK,\JKSone,\JKStwo].
Although the connection to the traditional TBA
is obscure, this formulation is more efficient in
numerical investigations for physical quantities  at finite temperatures.
For the $t-J$ model for instance, we need only three auxiliary
functions for a closed set of non-linear
integral equations. This results in
a numerical accuracy better than $10^{-6}$ [\JK,\JKSone]
for the specific heat, compressibility and other thermodynamic quantities.
Now the evaluation of correlation lengths is under
investigation.
We hope to report on explicit results based on
this formulation in the near future.

\beginsection{Acknowledgments}

The    authors   acknowledge   financial    support by  the   Deutsche
Forschungsgemeinschaft under grant  No.~Kl~645/3-1 and by the research
program       of       the          Sonderforschungsbereich       341,
K\"{o}ln-Aachen-J\"{u}lich. 

They would like to thank Z.  Tsuboi for discussions and collaboration at
the early stage of the present work. 
Furthermore, J.S. gratefully acknowledges the hospitality of the
Institut f\"ur Theoretische Physik der Universit\"at zu K\"oln.


\beginsection{References}

\item{[\YY]} C.N. Yang and C.P. Yang, J. Math. Phys. {\bf 10} (1969) 1115.
\item{[\Gaud]} M. Gaudin, Phys. Rev. Lett. {\bf 26} (1971) 1301. 
\item{[\HeisenTak]} M.~Takahashi, Prog. Theor. Phys. {\bf 46 }(1971) 401. 
\item{[\TakSuz]}   M.Takahashi and M. Suzuki, Prog. Theor. Phys. {\bf 48}
                (1972) 2187.
\item{[\Vla]}
A.A. Vladimirov, Non-String Two-Magnon Configurations in the Isotropic
  Heisenberg Magnet, Technical Report P17-84-409, Joint Institute for Nuclear
  Research Dubna,  (1984).
\item{[\KluZitt]} A.~Kl\"{u}mper and J.~Zittartz, Z. Phys. {\bf B 75 }(1989) 371.
\item{[\EKSstring]}  H.L. E{\ss}ler, V.E. Korepin and K.~Schoutens,
              J. Phys. {\bf A25} (1992) 4115.
\item{[\JuDo]} G.~J{\"u}ttner and B.D. D{\"o}rfel,
           J. Phys. {\bf A 26} (1993) 3105.
\item{[\spinkb]}  A. Kl{\"u}mper and M.T. Batchelor,
                  J. Phys. {\bf A 23} (1990) L189. 
\item{[\spinkbt]}  A. Kl{\"u}mper, M.T. Batchelor and P.A. Pearce,
                  J. Phys. {\bf A 24} (1991) 3111. 
\item{[\AlczMart]}  F.C. Alcaraz and M.J. Martins, J. Phys. {\bf A 21} 
                  (1988) L381,
                   ibid. 4397.
\item{[\AlZ]}    Al.B. Zamolodchikov, Phys. Lett. {\bf B253}, (1991) 391.  
\item{[\TsvBab]} H.M. Babujian  and A.M. Tsvelick, 
                 Nucl. Phys. {\bf B265} (1986) 24.
\item{[\MSuz]}   M. Suzuki, Phys. Rev. {\bf B 31} (1985) 2957.
\item{[\InSuz]} M.~Suzuki and M.~Inoue, Prog. Theor. Phys. {\bf 78} (1987) 787.
\item{[\Koma]}    T.~Koma, Prog. Theor. Phys. {\bf 78} (1987) 1213.
\item{[\SAW]}    J.~Suzuki, Y.~Akutsu and M.~Wadati, 
              J. Phys. Soc. Japan {\bf 59} (1990) 2667.
\item{[\SNW]}     J. Suzuki, T. Nagao and M. Wadati, 
          Int. J. Mod. Phys. {\bf B6} (1992) 1119-1180.
\item{[\TakQT]} M. Takahashi, Phys. Rev. {\bf B 43} (1991) 5788.
\item{[\Klu]}    A.~Kl\"umper, Ann. Physik {\bf 1 }(1992) 540.
\item{[\KZeit]}   A.~Kl\"umper, Z. Phys. {\bf B 91} (1993) 507.
\item{[\DdV]}    C.~Destri and H.J. de~Vega, Phys. Rev. Lett. 
                 {\bf 69} (1992) 2313.
\item{[\Mizu]}   H.~Mizuta, T.~Nagao and M.~Wadati,
                J. Phys. Soc. Japan {\bf 63} (1994) 3951.
\item{[\KWZ]} A. Kl{\"u}mper, T. Wehner and J. Zittarz, 
            J. Phys. A. {\bf 30} (1997) 1897.
\item{[\JK]} G. J{\"u}ttner and  A. Kl{\"u}mper, Euro. Phys. Lett. 
             {\bf 37}(1997) 335.
\item{[\JKSone]} G. J{\"u}ttner, A. Kl{\"u}mper and J. Suzuki,
               Nucl. Phys. {\bf B487} (1997) 650.
\item{[\JKStwo]} G. J{\"u}ttner, A. Kl{\"u}mper and J. Suzuki,
               J. Phys. A. {\bf 30} (1997) 1881.
\item{[\Cherednik]} I. Cherednik, in Proc.\ of the XVII International
Conference on Differential Geometric Methods in Theoretical Physics, Chester,
ed.\ A.I.\ Solomon, (World Scientific, Singapore, 1989)
\item{[\CPo]} V. Chari and A. Pressley,
          L'Enseignement Math. {\bf 36} (1990) 267.
\item{[\CPt]} V. Chari and A. Pressley, Comm. Math. Phys. {\bf 142} 
               (1991) 261. 
\item{[\KNS]} A. Kuniba, T. Nakanishi and J. Suzuki, 
         Int.J.Mod.Phys {\bf A9} (1994) 5215, ibid 5267.
\item{[\KN]}  A. Kuniba and T. Nakanishi, Mod. Phys. Lett. 
        {\bf A7} (1992) 3487.
\item{[\Martins]} M.J. Martins, Phys. Rev. Lett. {\bf 67} (1991) 419.
\item{[\Fendley]} P. Fendley, Nucl. Phys. {\bf B374} (1992) 667.
\item{[\KM]} T.R. Klassen, E. Meltzer, Nucl. Phys. {\bf B370} (1992) 511.
\item{[\KNSCh]} A. Kuniba, T. Nakanishi and J. Suzuki, 
     Mod. Phys. Lett. {\bf A8} (1993) 1649-1659.
\item{[\TateoDorey]} P. Dorey and R. Tateo, Nucl. Phys. {\bf B482} (1997) 639.
\item{[\BLZ]} V.V. Bazhanov, S.L. Lukyanov and A.B. Zamolodchikov,
            Nucl. Phys. {\bf B489} (1997) 487.
\item{[\Tsuboi]} Z. Tsuboi, preprint(1997) UT-KOMABA 97-8.
\item{[\Schlottmann]} P.~Schlottmann, Phys. Rev. {\bf B 36} (1987) 5177.
\item{[\EK]} F.H.L. E{\ss}ler and V.E. Korepin, condmat/9307019.
\item{[\EKS]} F.H.L. E{\ss}ler, V.E. Korepin and K. Schoutens, Phys. Rev. Lett.
            {\bf 68} (1992) 2960.
\item{[\SuthTJ]} B.~Sutherland, Phys. Rev. {\bf B 12} (1975) 3795.
\item{[\BB]} P.A. Bares, G.~Blatter, Phys. Rev. Lett. {\bf 64} (1990) 2567.
\item{[\BBO]} P.A. Bares, G.~Blatter and M.~Ogata,
            Phys. Rev. {\bf B 44} (1991) 130.
\item{[\Yang]}{C.N. Yang} Phys. Rev. Lett. {\bf 63} (1989) 2144.
\item{[\AS]} J.~de Boer, V.E.~Korepin, A.~Schadschneider: Phys. Rev. Lett. 
{\bf 74}, 789 (1995);\hfill\break
\hbox{{A. Schadschneider}: Phase Transitions {\bf 57} 37 (1996).}
\item{[\PS]} J.H.H. Perk and C.~Schulz, Phys. Lett. {\bf A 84} (1981) 407.
\item{[\Schulz]} C.L. Schulz, Physica {\bf A 122} (1983) 71.
\item{[\JSuz]} J.~Suzuki, J. Phys. {\bf A 25} (1992) 1769.
\item{[\RB]}R.Z. Bariev, J. Phys. {\bf A 27} (1994) 3381.
\item{[\Kl]} A. Kl\"umper, Int. J. Mod. Phys. B {\bf 11} 141 (1997)
\item{[\SuzGtwo]}  J.~Suzuki,  Phys. Lett. {\bf A195} (1994) 190.
\item{[\KS]}  A. Kuniba and J. Suzuki,
      Comm. Math. Phys. {\bf 173} (1995) 225-264.
\item{[\KOS]} A. Kuniba, Y. Ohta and  J. Suzuki, (1995)  J.Phys A {\bf 28}
        6211-6226.
\item{[\BazR]} V.V. Bazhanov and N.Yu Reshetikhin, 
             Int.J.Mod.Phys {\bf A4} (1989) 115.
\item{[\KP]} A. Kl{\"u}mper and P.A. Pearce, Physica {\bf A183} (1992) 304.
\item{[\Martinss]} For the representation theoretical meaning of this relation,
  see, T. Deguchi and P.P. Martin, Int. J. Mod. Phys. {\bf A7} (1992) 1A 165
  and  P.P. Martin and V. Rittenberg, 
  Int. J. Mod. Phys. {\bf A7} (1992) 1B 707.
\item{[\ZB]} Y.-K. Zhou and M.T. Batchelor, Nucl. Phys. B 490 [FS] (1997) 576.
\item{[\Rava]} F. Ravanini, R. Tateo and A. Valleriani, 
             Phys. Lett. {\bf B293} (1992) 361.
\item{[\KPHX]} For similar calculations on
           finite size correction for row-to-row transfer matrix of
            the Hard Hexagon Model, see
 A. Kl{\"u}mper and P.A. Pearce, Phys. Rev. Lett. {\bf 66} (1991) 974,
 J. Stat. Phys. {\bf 64} (1991) 13.

\end